\documentclass{moriond}
\usepackage{amssymb}
\usepackage{lineno}

\def\be{\begin{equation}}
\def\ee{\end{equation}}
\def\bea{\begin{eqnarray}}
\def\eea{\end{eqnarray}}

\newcommand{\Photo}{}

\begin{document}
\vspace*{0.8cm}
\title{Contribution to the 2026 Cosmology session of the 60th Rencontres de Moriond:\\
Galactic moments: Understanding polarized foregrounds complexity in the quest for CMB primordial $B$ modes with \textit{LiteBIRD}}



\author{S. Vinzl,$^1$ J. Aumont,$^1$ L. Vacher,$^2$ R. T. Génova-Santos,$^{3,4}$ D. Adak,$^3$ and A. Rizzieri$^5$\\
for the \textit{LiteBIRD} Collaboration}

\address{$^1$Univ Toulouse, CNES, CNRS, IRAP, Toulouse, France\\
$^2$Université Paris-Saclay, CNRS/IN2P3, IJCLab, 91405 Orsay, France\\
$^3$Instituto de Astrof\'isica de Canarias, E-38200 La Laguna, Tenerife, Canary Islands, Spain\\
$^4$Departamento de Astrof\'isica, Universidad de La Laguna (ULL), E-38206, La Laguna, Tenerife, Spain\\
$^5$Department of Physics, University of Oxford, Denys Wilkinson Building, Keble Road, Oxford OX1 3RH, United Kingdom}

\maketitle

\abstracts{
Accurate modeling of polarized Galactic emission has become a major challenge for current and next-generation cosmic microwave background (CMB) $B$-mode experiments. Ignoring the spectral complexity of thermal dust and Galactic synchrotron emission when integrating along the line of sight and over large sky fractions inevitably leads to biases in CMB polarization analyses. In this work, we review how the future \textit{LiteBIRD} satellite, which will benefit from an increased number of bands and sensitivity with respect to past CMB experiments such as \textit{Planck}, will exploit its broader frequency coverage to characterize the spectral properties of interstellar medium emission across the three-dimensional structure of the Milky Way. We show that the canonical description of foreground spectral behavior will reach its limits for \textit{LiteBIRD}, and that this challenge can be addressed using the moment expansion formalism.
}

\section{Introduction}

The main limiting factor for current and future cosmic microwave background (CMB) experiments is undeniably our ability to separate the faint CMB primordial signal from the emission originating from our Galaxy. In polarization, the latter originates from thermal dust emission which is dominant at frequencies $\gtrsim 100$~GHz, and synchrotron radiation at lower frequencies.

\textit{LiteBIRD} is part of the next-generation CMB $B$-mode experiments having the goal of putting tighter constraints on the tensor-to-scalar ratio $r$, quantifying the amplitude of primordial gravitational waves generated during inflation. In this work,~\cite{Vinzl2026} we investigate its ability to improve our understanding of diffuse Galactic polarized emission with respect to \textit{Planck}. \par 

We simulate maps at \textit{LiteBIRD} frequencies in the range $40$-$402$~GHz using the corresponding noise levels and beam widths.~\cite{PTEP2023} We consider three Galactic emission complexities:~\cite{Panexp2025} {\tt low} with spatially constant spectral parameters, {\tt medium} with spatial variations orthogonal to the line of sight, and {\tt high} with three-dimensional variations of the dust emission properties. Cutting low Galactic latitudes using a sky fraction $f_{\rm sky} = 0.7$, we compute $B$-mode cross-frequency angular power spectra $\mathcal{C}_\ell^{\nu_i \times \nu_j}$. We fit them by modeling the dust spectral energy distribution (SED) as a modified black-body (MBB) with amplitude $A_{\rm d}$, spectral index $\beta_{\rm d}$ and temperature $T_{\rm d}$, and synchrotron as a power law (PL) with amplitude $A_{\rm s}$ and index $\beta_{\rm s}$. The spatial correlation between the two components is quantified by the correlation coefficient $\rho$. We model the same spectra using the moment expansion formalism, in which variations of the emission properties are taken into account by Taylor-expanding the SED around pivot values of the spectral parameters.~\cite{Chluba2017}$^,$~\cite{Vacher2022a} 

\section{\boldmath Constraining $B$-mode Galactic emission with \textit{LiteBIRD}}

\begin{figure}[t]
    \centering
    \includegraphics[height=72mm]{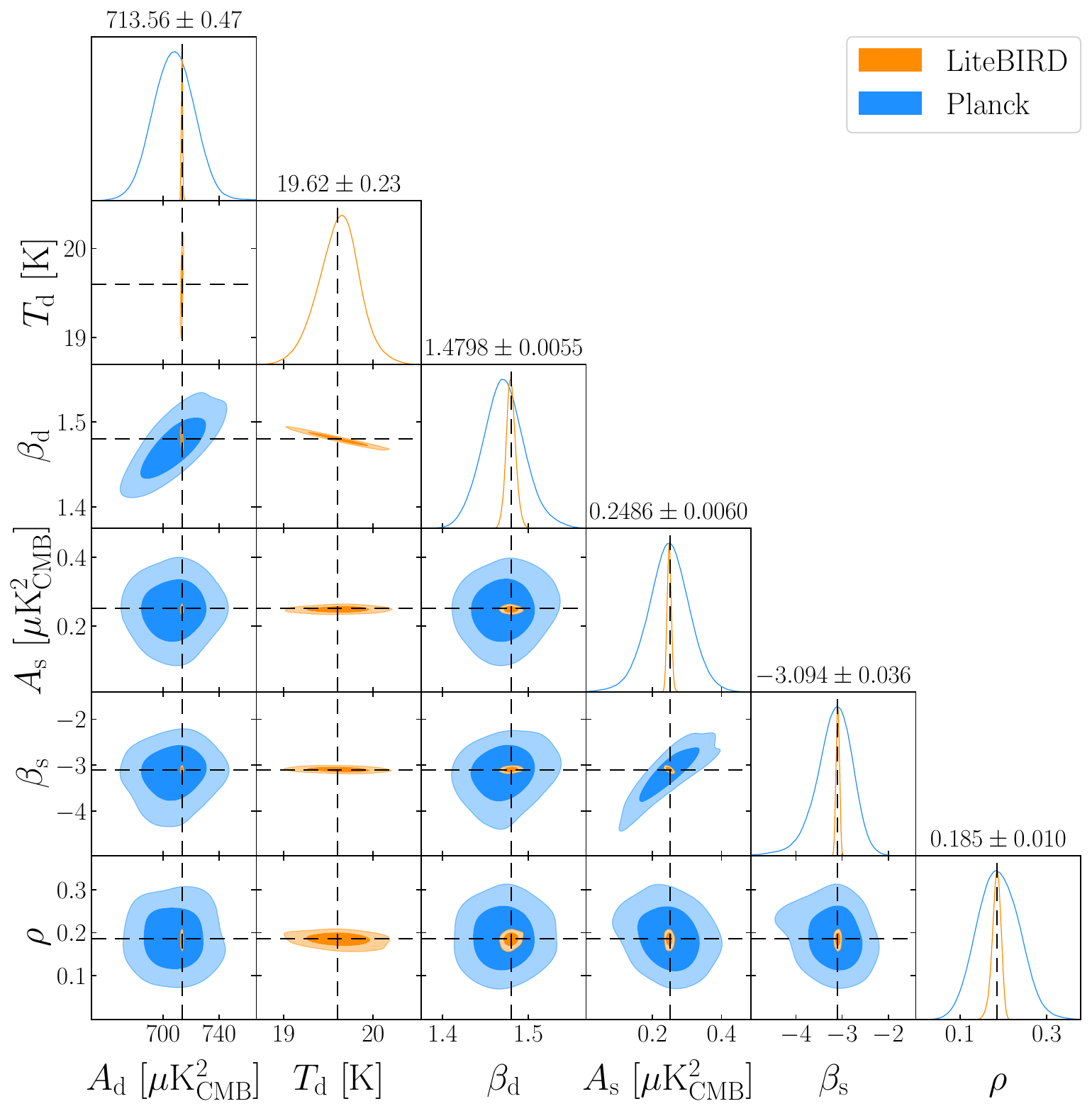}
    \includegraphics[height=72mm]{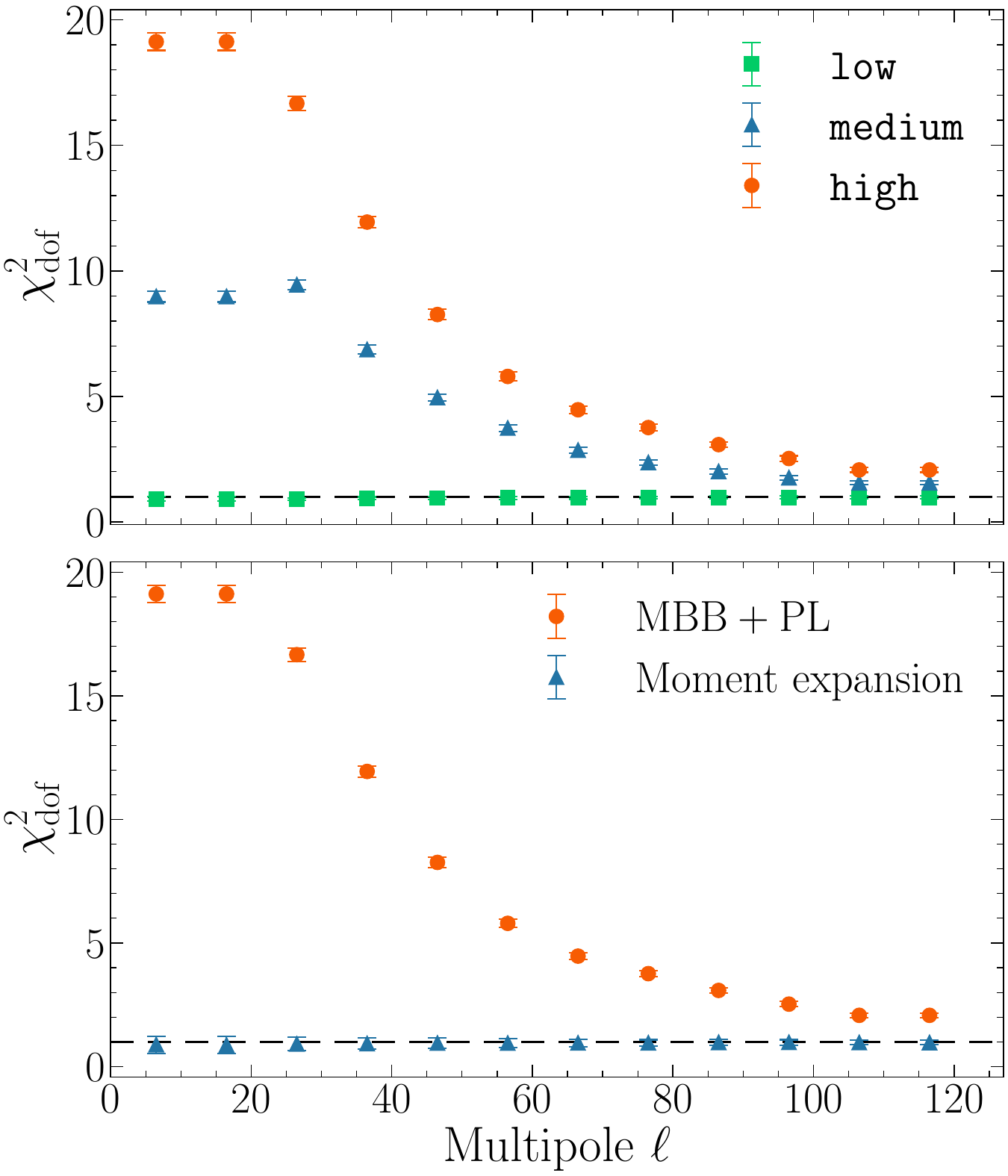}
    \caption{Left: Posterior distribution of MBB+PL parameters obtained with \textit{LiteBIRD} and \textit{Planck} {\tt low} simulations at $\ell = 22 \rightarrow 31$. Dashed lines show the input values. Right: Reduced $\chi^2$ obtained from the fits to \textit{LiteBIRD} simulations (top). Adding moments to the model decreases $\chi^2_{\rm dof}$ for {\tt high} complexity simulations (bottom).}
    
    \label{fig:figure}
\end{figure}

The left panel of Fig.~\ref{fig:figure} shows the posterior distribution obtained for the MBB+PL model at multipoles $\ell = 22 \rightarrow 31$ for {\tt low} complexity, comparing \textit{LiteBIRD} and \textit{Planck}. We clearly see the improvements that will be brought by \textit{LiteBIRD}, enabling tighter constraints on all parameters.

While the reduced $\chi^2$ is in this case consistent with one at all scales, the simple MBB+PL model does not manage to capture the full complexity of the sky for {\tt medium} and {\tt high}, as it is shown in the upper right panel. Indeed, $\chi^2_{\rm dof}$ is strongly increased at largest scales, highlighting the variations of the spectral parameters across the Galaxy. The lower panel shows that moment expansion enables a satisfactory modeling of the SED distortions caused by these variations. 

\section{Conclusion}

\textit{LiteBIRD} will be able to accurately constrain the spectral properties of Galactic polarized foregrounds and to detect their variations across the three dimensions of the Milky Way. By doing so, it will rule out the common MBB+PL parametrization, requiring a precise modeling of the SED distortions due to the averaging of the emission along and between lines of sight. The proper framework is given by the moment expansion formalism, which will play a crucial role in parametric component separation methods in the quest for CMB primordial $B$ modes.

\section*{Acknowledgments}

%
\textit{LiteBIRD} (phase A) activities are supported by the following funding sources: ISAS/JAXA, MEXT, JSPS, KEK (Japan); CSA (Canada); CNES, CNRS, CEA (France);
DFG (Germany); ASI, INFN, INAF (Italy); RCN (Norway); MCIN/AEI, CDTI (Spain); SNSA, SRC (Sweden); UKSA (UK); and NASA, DOE (USA).
%

\section*{References}
\bibliography{biblio}

@ARTICLE{Vinzl2026,
       author = {{Vinzl, S.} et al.},
      journal = {arXiv e-prints},
         year = 2026,
        month = jul,
          eid = {arXiv:2607.20080},
        pages = {arXiv:2607.20080},
archivePrefix = {arXiv},
       eprint = {2607.20080},
 primaryClass = {astro-ph.GA},
       adsurl = {https://ui.adsabs.harvard.edu/abs/2026arXiv260720080V}
}

@ARTICLE{PTEP2023,
       author = {{LiteBIRD Collaboration}},
      journal = {PTEP},
         year = 2023,
        month = apr,
       volume = {2023},
       number = {4},
          eid = {042F01},
        pages = {042F01},
          doi = {10.1093/ptep/ptac150},
archivePrefix = {arXiv},
       eprint = {2202.02773},
 primaryClass = {astro-ph.IM},
       adsurl = {https://ui.adsabs.harvard.edu/abs/2023PTEP.2023d2F01L}
}

@ARTICLE{Panexp2025,
       author = {{Pan-Experiment Galactic Science Group}},
      journal = {\apj},
         year = 2025,
        month = sep,
       volume = {991},
       number = {1},
          eid = {23},
        pages = {23},
          doi = {10.3847/1538-4357/adf212},
archivePrefix = {arXiv},
       eprint = {2502.20452},
 primaryClass = {astro-ph.CO},
       adsurl = {https://ui.adsabs.harvard.edu/abs/2025ApJ...991...23P}
}

@ARTICLE{Chluba2017,
       author = {{Chluba, J.} et al.},
      journal = {\mnras},
         year = 2017,
        month = nov,
       volume = {472},
       number = {1},
        pages = {1195-1213},
          doi = {10.1093/mnras/stx1982},
archivePrefix = {arXiv},
       eprint = {1701.00274},
 primaryClass = {astro-ph.CO},
       adsurl = {https://ui.adsabs.harvard.edu/abs/2017MNRAS.472.1195C}
}

@ARTICLE{Vacher2022a,
       author = {{Vacher, L.} et al.},
      journal = {\aap},
         year = 2022,
        month = apr,
       volume = {660},
          eid = {A111},
        pages = {A111},
          doi = {10.1051/0004-6361/202142664},
archivePrefix = {arXiv},
       eprint = {2111.07742},
 primaryClass = {astro-ph.CO},
       adsurl = {https://ui.adsabs.harvard.edu/abs/2022A&A...660A.111V}
}


\end{document}